\documentclass{ws-mpla}

\begin{document}

\def\fdg{\hbox{$.\!\!^\circ$}}
\def\farcm{\hbox{$.\mkern-4mu^\prime$}}
\def\farcs{\hbox{$.\!\!^{\prime\prime}$}}

\markboth{Umetsu et al.}
{Simulation of a combined SZE and weak lensing cluster survey for AMiBA experiment}

%
\catchline{}{}{}{}{}
%

\title{SIMULATION OF A COMBINED SZE AND WEAK LENSING CLUSTER SURVEY
 FOR AMIBA EXPERIMENT
}

\author{\footnotesize KEIICHI UMETSU
}

\address{Institute of Astronomy and Astrophysics, Academia Sinica,
P.O. Box 23-141, Taipei 106, Taiwan, Republic of China\\
keiichi@asiaa.sinica.edu.tw}

\author{TZIHONG CHIUEH, KAI-YANG LIN, JUN-MEIN WU}
\address{Department of Physics, National Taiwan University, 
Taipei 106, Taiwan, Republic of China
}

\author{YAO-HUAN TSENG}
\address{Institute of Astronomy and Astrophysics, Academia Sinica,
P.O. Box 23-141, Taipei 106, Taiwan, Republic of China\\
}

\maketitle

\pub{Received (Day Month Year)}{Revised (Day Month Year)}

\begin{abstract}

We present simulations of 
interferometric 
Sunyaev-Zel'dovich effect (SZE) and
optical weak lenisng observations
for the forthcoming AMiBA experiment,
aiming at searching for high-redshift clusters of galaxies.
On the basis of simulated sky maps,
we have derived 
theoretical halo number counts and redshift distributions of
selected halo samples 
for an AMiBA SZE survey and a 
weak lensing follow-up  survey.
By utilizing the conditional number counts of weak lensing halos
with the faint SZE detection, we show that a combined SZE and
weak lensing survey can 
gain an additional fainter halo sample 
at a given false positive rate,
which cannot be obtained from either survey alone.



%

\end{abstract}

\ccode{PACS Nos.: include PACS Nos.}

\section{Introduction}	
The thermal Sunyaev-Zel'dovich effect (SZE\cite{Birkinshaw}) is a spectral
distortion
of the Cosmic Microwave Background (CMB) radiation due to 
the inverse-Compton scattering of CMB photons by high energy 
electrons in the intracluster medium (ICM). 
The most remarkable properties of the SZE is that its surface brightness is 
redshift independent, which make it as an ideal probe of the
high-redshift universe. Further, since the SZE is proportional to
the thermal energy content of the ICM,
SZE imaging surveys allow us to 
select clusters over a wide range of the redshift
with physically meaningful selection criteria. Weak gravitational
lensing, on the other hand, probes the total mass projected along
the line-of-sight, and hence 
provides complementary information 
on the mass of galaxy clusters.\cite{Bartelmann} 
Array for Microwave Background Anisotropy (AMiBA\cite{AMiBA})
is a 19-element interferometric array with full polarization
capabilities operating at $95$GHz, specifically designed for the CMB
observations. One of the main scientific goals of AMiBA is to conduct
blind SZE surveys 
to search for high-redshift clusters.
AMiBA will also conduct 
follow-up optical imaging observations 
with wide-field camera, MegaCam, at {\it Canada France Hawaii Telescope}
(CFHT). 
In this paper, we simulate
the forthcoming AMiBA SZE experiment combined with the planned follow-up
weak lensing observations to examine the expected cluster number counts
%
for individual surveys, and explore the potential of a combined 
AMiBA SZE and weak lensing cluser survey.
\section{Simulation Data and Mock Observations}
%
%

To make sky maps with realistic SZE and weak lensing signals,
we use results from preheating cosmological simulations
of a $\Lambda$CDM model 
$(\Omega_m=0.34,\Omega_{\Lambda}=0.66,\Omega_b=0.044,h=0.7,\sigma_8=0.94)$
in a $100h^{-1}$Mpc co-moving box 
which reproduce the observed cluster $M_X$-$T_X$
and $L_X$-$T_X$ relations at $z=0$.\cite{Kyle}
We construct 36 SZE sky maps
each with $1$deg$^2$ on a $1024^2$ grid, by projecting the electron
pressure through the randomly displaced and oriented simulation boxes,
separated by $100h^{-1}$ Mpc, along a viewing cone out  to the redshift
of $z=2$. Similarly, weak lensing convergence ($\kappa$)
maps are constructed by projecting
the distance-weighted mass over-density $\delta\rho$
out to a source plane at $z=z_s$. 

As an AMiBA specification,
we adopt a close-packed hexagonal configuration of $19\times 1.2$m dishes
on a single platform.
This array configuration yields a synthesized beam of ${\rm
FWHM}\simeq 2'$ ($0.6h^{-1}$Mpc
at $z=0.8$),
which is  optimized to detect high-redshift clusters. The field-of-view
of the primary beam is about $10'$.
To generate the mock AMiBA visibility data, 
we perform a mosaic survey of  $9\times 9$ pointings with 
each exposure of $t_{\rm exp}=1.5$ hours 
covering 
$1$deg$^2$ with spacing of  $4\farcm 5$.
%
The AMiBA sensitivity is then given as $\sigma_{\rm noise}^S=
0.78(c_{\rm mos}/0.83)(\eta/0.6)^{-1}(T_{\rm sys}/70{\rm K})
(n_{\rm pol}/2)^{-1/2}(B/20{\rm GHz})^{-1/2}(t_{\rm exp}/1.5{\rm
hours})^{-1/2}$ mJy/beam,
where $c_{\rm mos}$ depends on the mosaicking strategy.
The non-linear maximum entropy method is
applied to reconstruct an AMiBA SZE map 
from the mock visibility data.
For a weak lensing survey,
we assume moderately deep optical observations with 
a mean source number density of $\bar{n}=30{\rm arcmin}^{-2}$ and 
$z_s=1$.
We take into account noise in observable image ellipticities
due to the random-phase intrinsic source ellipticities.
We choose the intrinsic dispersion of $\sigma_{\rm int}=0.4$.
We then apply a pixelization on the mock weak lensing data using
a Gaussian filter with ${\rm FWHM}=2'$, and perform a linear mass
inversion to reconstruct the $\kappa$-map.
The sensitivity in reconstructed $\kappa$-maps is given as
$\sigma_{\rm noise}^{\kappa}=1.75\times 10^{-2}(\sigma_{\rm int}/0.4)
(\bar{n}/30{\rm arcmin}^{-2})^{-1/2}({\rm FWHM}/2')^{-1}$.
We carry out the statistical analysis on reconstructed sky maps
by using the inner $41' \times 41'{\rm arcmin}^2$ subfield 
in order to avoid the noisier boundaries. Thus the effective survey
area used for our statistical analysis is $36\times 0.47{\rm deg}^2
\sim 17{\rm deg^2}$.

\section{Results}

Figure 1 shows the theoretical halo number counts $N(>\nu)$
and false positive rate
as a function of peak threshold $\nu(=S/\sigma_{\rm noise}^S,
\kappa/\sigma_{\rm noise}^{\kappa})$ 
obtained from mock AMiBA SZE and weak
lensing surveys. To quantify the spurious peak detections
 due to experimental noise, 
we followed the prescription given by Ref.~\refcite{Pen}.
The limiting halo masses $M_{\rm lim}(z)$
for both surveys are shown in Figure 2.
For both surveys, halo samples are 
defined at a false positive rate of $10\%$, and the 
redshift distributions of the halo samples are derived (see Figure 2).
Further, we utilize the conditional number counts of weak lensing
halos with faint SZE detection, $N(\kappa>\kappa_{\rm lim}
|S>S_{\rm lim})$,
to explore the potential of  a combined AMiBA SZE and weak lensing
survey. Here we choose the SZE peak threshold of
$S_{\lim}=2.5\sigma_{\rm noise}^S=2{\rm mJy/beam}$ (see Figure 1).
Table 1 summarizes the main properties of halos samples obtained from
mock AMiBA SZE and weak lensing observations.

\section{Conclusions}
We have examined the expected cluster number counts 
and redshift distributions of cluster samples
for the forthcoming
AMiBA SZE/weak lensing cluster survey on the basis of 
$\Lambda$CDM cosmological simulations.
By utilizing the conditional halo number counts 
$N(\kappa>\kappa_{\rm lim}
|S>S_{\rm lim})$,
we have demonstrated that a combined SZE and
weak lensing survey can 
gain an additional fainter halo sample 
at a given false positive rate,
which cannot be obtained from either survey alone.

\section*{Acknowledgments}
We would like to thank Mark Birkinshaw for fruitful discussions.
AMiBA is funded by the Ministry of Education and National Science
Council in Taiwan.

\clearpage

\begin{figure}[th]
\centerline{\psfig{file=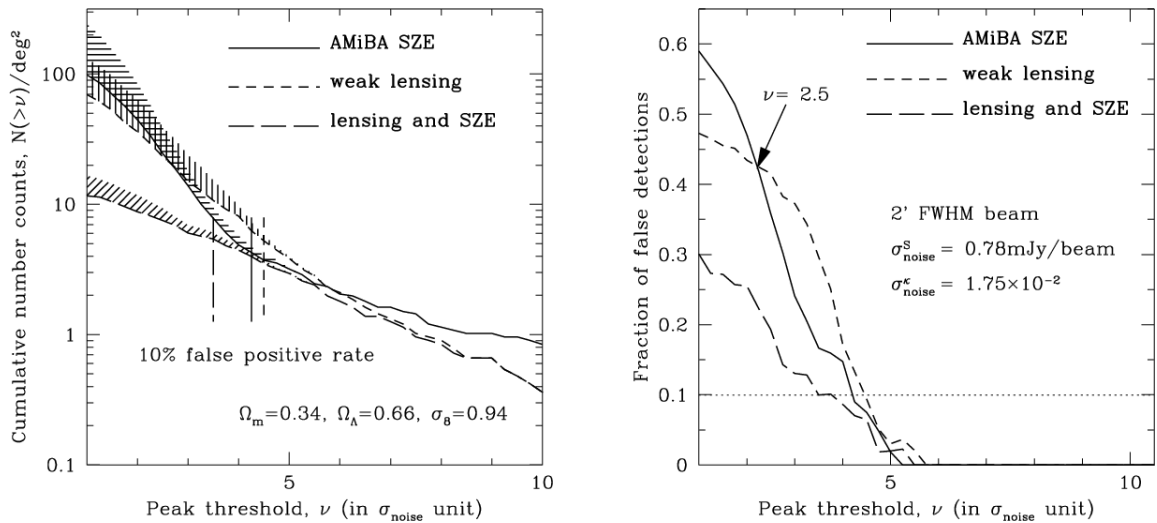,width=6in}}
\vspace*{8pt}
\caption{Left: Cumulative distribution functions $N(>\nu)$ 
of halo number counts
in mock AMiBA (solid)
and weak lensing (dashed)
observations as a function of peak threshold $\nu
(=S/\sigma_{\rm noise}^S, \kappa/\sigma_{\rm noise}^{\kappa})$
derived from 36 simulated sky maps ($17{\rm deg}^2$). 
The long-dashed curve represents
the  conditional number counts of weak lensing halos 
with faint SZE detection of $S>2.5\sigma_{\rm noise}^{S}=2$mJy/beam. The shaded regions indicate
false positive detections due to noise. 
Right: fraction of false detections  as a function of peak threshold.
}
\end{figure}

\clearpage

\begin{figure}[th]
\centerline{\psfig{file=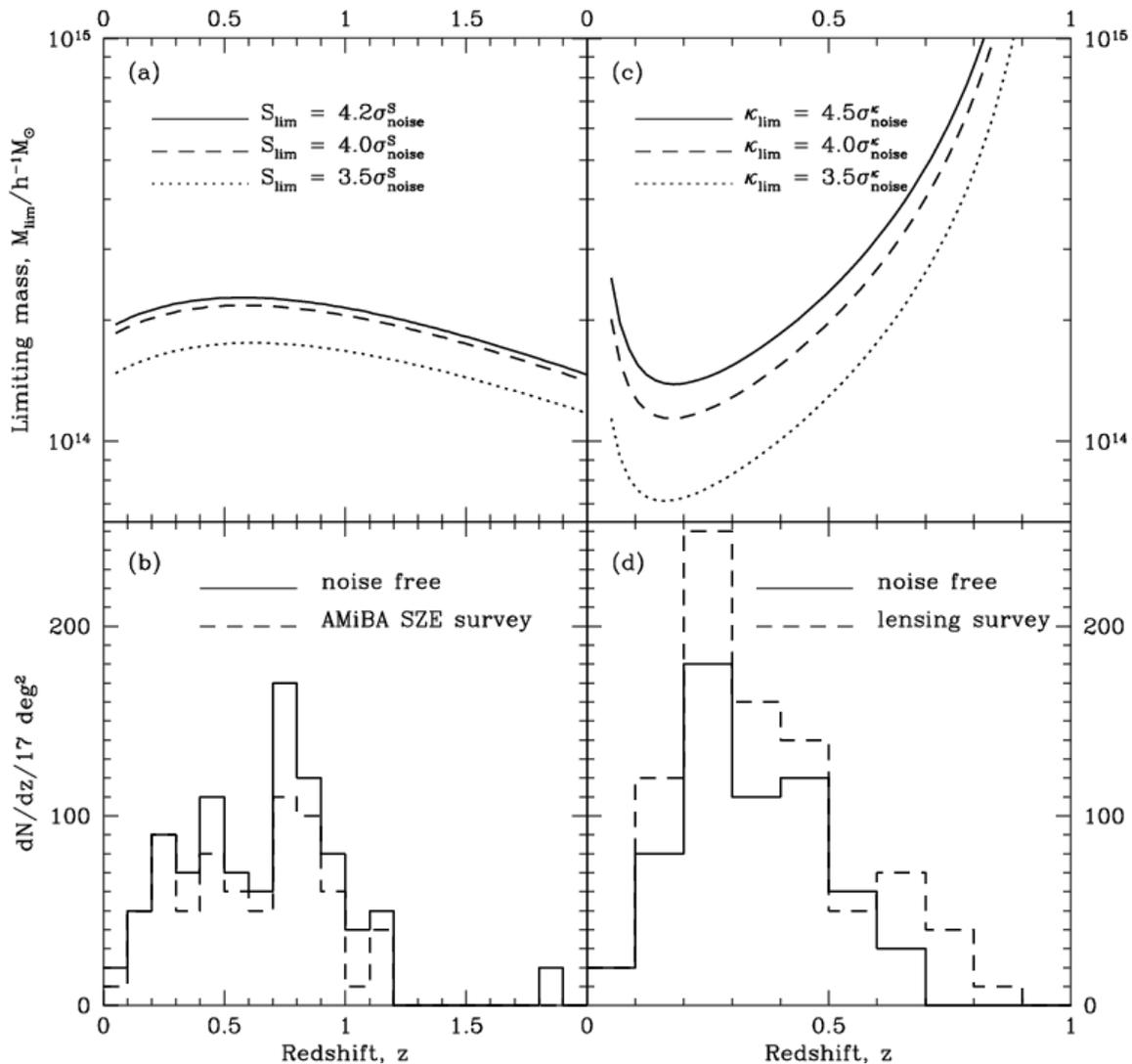,width=6in}}
\vspace*{8pt}
\caption{
Top: the limiting mass $M_{\rm lim}$ of halos as a function
of redshift $z$ for (a) AMiBA SZE and (c) weak lensing surveys
(see Section 2). $M_{\rm lim}(z)$ is calculated for three different
peak thresholds. The solid curves correspond to a false positive
rate of $10\%$. Bottom: redshift distributions of halo samples
selected from mock $17{\rm deg}^2$ 
(b) AMiBA SZE and (d) weak lensing surveys.
}
\end{figure}

\clearpage

\begin{table}[h]
\tbl{Halo samples selected from mock AMiBA SZE and weak lensing surveys}
{\begin{tabular}{@{}lccc@{}} \toprule
Halo sample & limiting peak-flux $S$ & limiting peak $\kappa$-value & $N$ (deg$^{-2}$)\\
       & $\nu = S/\sigma_{\rm noise}^S$ & $\nu=\kappa/\sigma_{\rm
  noise}^{\kappa}$& \\ \colrule
AMiBA SZE    & $\nu >4.2$ &  & 4.2 \\
weak lensing &            & $\nu>4.5$ & 5.2 \\
combined AMiBA/lensing & $\nu>2.5$           & $\nu>3.5$ & 4.9 \\
faint & $2.5 < \nu <4.2$  & $3.5<\nu<4.5$  & 1.4 \\ \botrule
\end{tabular}}
\end{table}


\section*{References}

\vspace*{6pt}

\begin{thebibliography}{0}


\bibitem{Birkinshaw} M. Birkinshaw, {\it Phys. Rep.}, {\bf 310}, 97 (1999).

\bibitem{Bartelmann} M. Bartelmann and P. Schneider, {\it
	Astron. Astrophys.}, {\bf 345}, 17

\bibitem{AMiBA} 
 K. Y. Lo, T. Chiueh, H. Liang, C.-P. Ma, R. N. Martin, K.-W. Ng,
 U.-L. Pen, and R. Subramanyan, {\it in IAU Symp. 201, New
	Cosmological Data and the Values of the Fundamental Parameters,
	ed. A. Lasenby and A. Wilkinson (San Francisco: ASP)}, 31 (2000).

\bibitem{Kyle}
K.-Y. Lin, L. Lin, T.-P. Woo, Y.-H. Tseng, and T. Chiueh,
{\it astro-ph}/0210323


\bibitem{Pen}
P. Zhang, U.-L. Pen, and B. Wang, {\it Astrophys. J.}, {\bf 577}, 555


\end{thebibliography}
\end{document}